\documentclass[iop]{emulateapj-rtx4}

\slugcomment{Draft Version, Accepted for Publication in ApJL}

\shorttitle{Gradual Acceleration of M 87 Jet}
\shortauthors{Asada, K. et al.}

\begin{document}


\title{Discovery of Sub- to Superluminal Motions in the M87 Jet: An Implication of  the  Acceleration from Sub-relativistic to Relativistic Speeds}


\author{Keiichi Asada\altaffilmark{1}, Masanori Nakamura\altaffilmark{1}}
\affil{
$^1$Institute of Astronomy \& Astrophysics, Academia Sinica, P.O. Box
23-141, Taipei 10617, Taiwan \\ asada@asiaa.sinica.edu.tw, nakamura@asiaa.sinica.edu.tw}

\author{Akihiro Doi\altaffilmark{2}}
\affil{Institute of Space and Astronautial Science, Japan Aerospace Exploration Agency, \\ 3-1-1 Yoshinodai, Chuou-ku, Sagamihara, Kanagawa, 229-8510, Japan}

\author{Hiroshi Nagai\altaffilmark{3}}
\affil{National Astronomical Observatory of Japan, 2-21-1, Osawa, Tokyo, 181-8588, Japan}

\and

\author{Makoto Inoue\altaffilmark{1}}
\affil{$^1$Institute of Astronomy \& Astrophysics, Academia Sinica, P.O. Box 23-141, Taipei 10617, Taiwan}

\begin{abstract}
The velocity field of the M87 jet from milli-arcsecond (mas) to arcsecond scales is extensively investigated together with new radio images taken by EVN observations. 
We detected proper motions of components located at between 160 mas from the core and  the HST-1 complex for the first time.  
Newly derived velocity fields exhibits a systematic increase from sub-to-superluminal speed in the upstream of HST-1. 
If we assume that the observed velocities reflect the bulk flow, we here suggest that the M87 jet may be gradually accelerated through a distance of 10$^{6}$ times of the Schwarzschild radius of the supermassive black hole. 
The acceleration zone is co-spatial with the jet parabolic region, which is interpreted as the collimation zone of the jet (Asada \& Nakamura). The acceleration and collimation take place simultaneously, which we suggest a characteristic
of magnetohydrodynamic flows.  Distribution of the velocity field has a peak at HST-1, which is considered as the site of over-collimation, and shows a deceleration downstream of HST-1 
where the jet is conical. Our interpretation of the velocity map in the M87 jet gives a hypothesis in AGNs that the acceleration and collimation zone of relativistic jets extends over the 
whole scale within the sphere of influence of the supermassive black hole. 
\end{abstract}





\begin{figure*}[htbp]
\epsscale{1.15}
\plotone{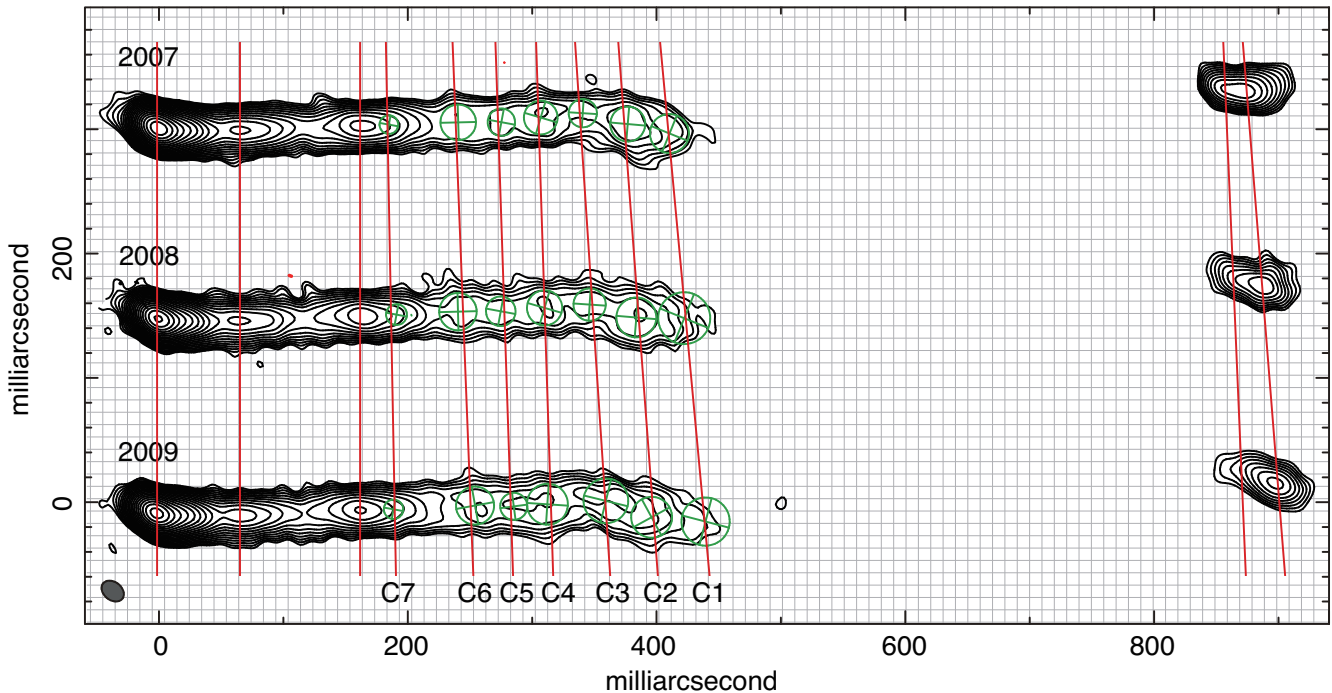}
\caption{ Contours are plotted at -1, 1, 1.4142, ..., 1024 $\times$ 2.12 mJy beam$^{-1}$, 
which is three-times the residual r.m.s. noise in the first epoch image.  
The synthesized beam is 19.9 mas $\times$ 14.6 mas with the major axis at a position angle of 73.4$^{\circ}$ 
for the first epoch.  Model components and trajectories are represented by green circles and red line, 
respectively.  
\label{IPOL-maps}}
\end{figure*}

\section{Introduction}

Understanding the acceleration mechanism of relativistic jets is one of the long--standing issues in high energy astrophysics.  
Observations of proper motion with a high angular resolution have been established as a useful tool to trace the kinematics of relativistic jets.  
For such proper motions, 
in the case of active galactic nuclei (AGN) jets, systematic studies have been conducted with the technique of very long baseline interferometry (VLBI).  
It has been reported that terminal velocities of the jets can reach up to Lorentz Factor ($\Gamma$) of $\sim$ 30 in VLBA 2 cm survey/MOJAVE samples 
\citep[]{K04, L09}.  However, observational samples to address how and where jets are accelerated are still insufficient.  

M 87 is the one of the closest AGNs \citep[$D$ = 16.7 Mpc; ][]{J05} having a relativistic jet.  
At this distance, one milli arcsecond (mas) corresponds to 0.084 pc, and 1 mas yr$^{-1}$ corresponds to 0.27 $c$.  
Thus, it enables us to measure wide range of the velocities from sub- to superluminal speed with current VLBI technique.  
In addition, since the mass of the central supermassive black hole is enormous \citep[{3.2  - 6.6 $\times$ 10$^{9}$ M$_{\sun}$;}][]{M97, G11, W05}\footnote{Throughout the paper, we adopt  black hole mass of 6.6 $\times$ 10$^{9}$ M$_{\sun}$.}, 
the Schwarzschild radius ($r_{s}$) is  131 AU = 0.0078 mas.  It provides a unique opportunity to explore the relativistic jet with the largest angular scale in terms of $r_{s}$. 

At arcsecond scales, a one-sided jet has been detected from radio through optical to X-ray \citep[e.g., ][]{O89, S96, P99, M02}.
At the mas scale, an existence of the counter jet is reported with VLBA observations at 15 and 43 GHz \citep[]{L04, K07, W08}, though the approaching side of the jet still dominates its emission.  
Very recently, the ``central engine'' of M 87, probably indicating the position of the black hole and accretion disk, is determined with phase-reference VLBI observations 
to be located at only 21 $\mu$as upstream from the VLBI core at 43 GHz in projected distance \citep[]{H11}.  

Proper motions of the M87 jet have been probed by VLBI for the innermost region ($<$ 1 arcsec from the core), and by the Very Large Array (VLA) and the Hubble Space Telescope (HST) for the outer region ($>$ 1 arcsec from the core) during the past two decades.  
VLBI observations at cm wavelength have been detected only subluminal ($<$ 0.3 $c$) or no proper motions within error \citep[e.g., ][]{R89, D06, K07}.  
The range of speeds were intensively measured with  VLBA  at 15 GHz, 
and it was revealed that only very slow proper motions between  0.003 $c$ to 0.05 $c$ exist in the range from 1 to 20 mas  from the nucleus \citep[]{K07}.   
On the other hand, recent VLBA observations at 43 GHz claimed to detect very fast proper motions; 1.1 $c$ or more at 0.5 mas \citep[]{A09} and say 2 $c$ in the range from 1 to 5 mas \citep[]{W08}.  
These observed speeds are different from the previous observations.  We will discuss this discrepancy in section 4.3.  

HST observations revealed trails of superluminal features at around 1 arcsecond from the core (a region named as HST-1) with a range of 4 $c$ - 6 $c$ \citep[]{B99}. 
VLA and HST observations detected both super- and subluminal motions downstream of HST-1 \citep[]{B95, MET13}.  
Both HST and VLA observations give a global trend that observed proper motions decrease as a function of the distance from HST-1 smoothly.

To summarize the observed proper motions, an interesting picture is derived so far; no evidence for highly relativistic velocities between the core and HST-1, 
and then, suddenly, superluminal motion of components immediately HST-1.  
There is a missing link on the velocity field of the M87 jet between 160 and 900 mas.  
Exploring the velocity field on this spatial scale is the key to investigate the dynamics of the M87 jet,  and provides us with clues to understand the acceleration mechanism of the jet. 


\section{Observation and Data Reduction}

We conducted monitoring observations of M 87 on 12 March 2007, 2 March 2008, and 7 March 2009 
using the EVN and Multi-Element Radio Linked Interferometer Network (MERLIN) at a wavelength of 18 cm. 
EVN observations were conducted with Cambridge (UK), Effelsberg (Germany), Jodrell Bank (UK), 
Knockin (UK: only at last epoch), Medicina (Italy), Noto (Italy), Onsala (Sweden), Torun (Poland), 
and Westerbork (Netherlands) stations.  Both left and right circular polarization data were recorded 
at each telescope using 8 channels of 8 MHz bandwidth and 2 bit sampling. The data were correlated 
at the Joint Institute for VLBI (JIVE) correlator.  

{\it A priori} amplitude calibration for each station was derived from a measurement of system 
temperatures during each run and the antenna gain.  Fringe fitting was performed using AIPS.  
After delay and rate solutions were determined, the data were averaged
over 12 seconds in each IF and self-calibrated using Difmap.

For the  self-calibrated images, we performed model fitting using Difmap.  
The core is defined by Gaussian model fitting for the innermost bright region.  
Both of the circular or elliptical Gaussian models have been applied, but the relative position 
of each component with respect to the core is well determined.  We summarize the results on the fitting in table 1.
After fitting, we cross-identified the components using not only their positions, but also their size 
and flux density.  Fitted size of the components agree from one epoch to next with assuming less than 10 \% error 
for each measurement in most cases.  Measured flux densities agree from one epoch to next with assuming 
less than 20 \% error for each measurementin most cases 
as well.  Therefore, we think  our identification is robust.

\begin{figure*}
\epsscale{1.15}
\plotone{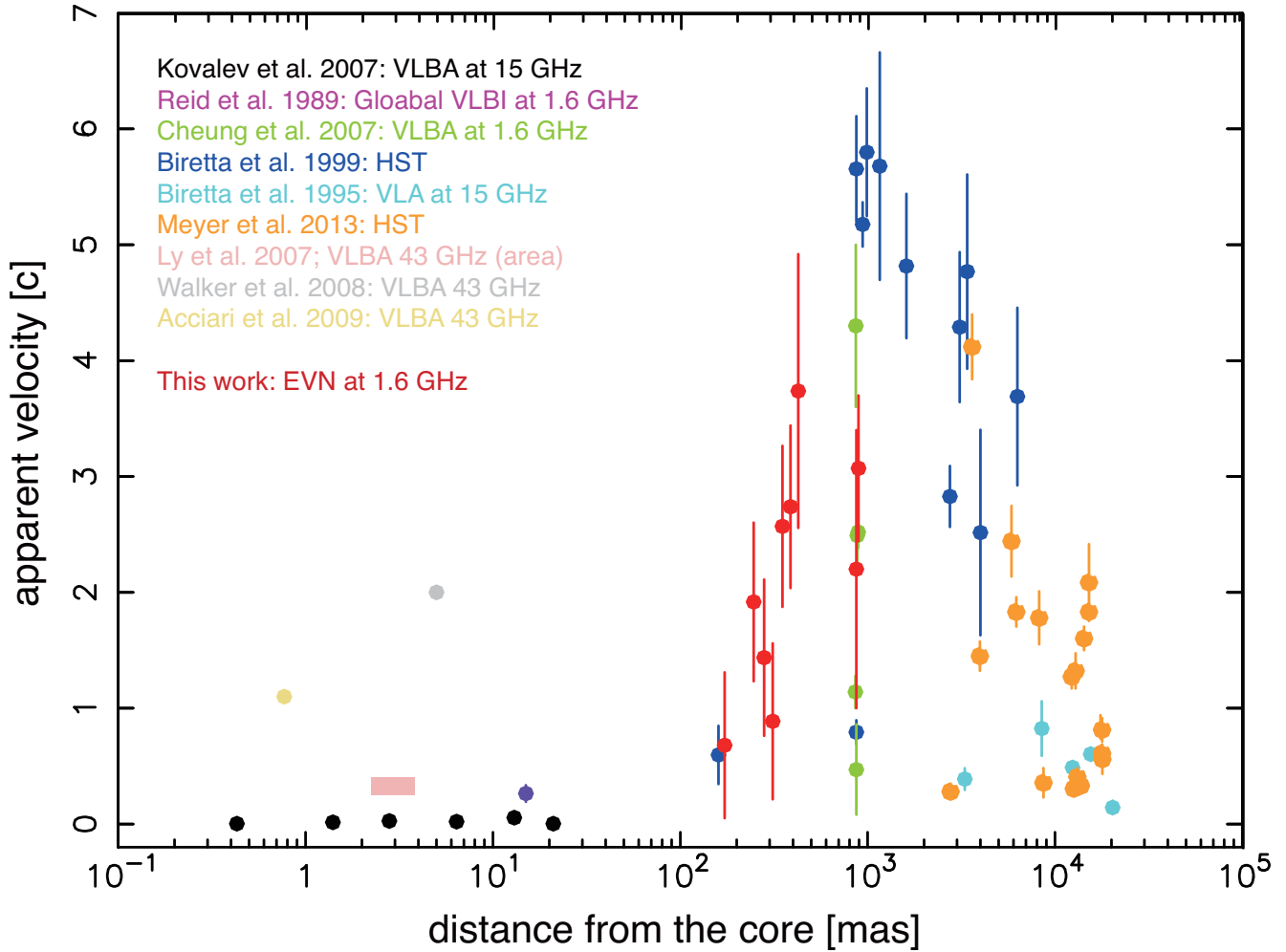}
\caption{ Distribution of the apparent velocity as a function of the projected distance from the core.  
Measurements by VLBA at 15 GHz (black circles), global VLBI at 1.6 GHz (purple circle), 
our measurements (red circles), VLBA at 1.6 GHz (green circles), HST (blue circles), 
VLA at 15 GHz (light blue circles).  Error bars represent $\pm$ 1 $\sigma$.  
EVN points connect the subluminal speed measure towards innermost jet with VLBA at 15 GHz and superluminal speed measured towards HST-1 with HST and VLBA.
We also plotted VLBA measurement at 43 GHz by Walker et al. (2008) and Acciari et al. (2009) as a reference.  
We also show the velocity measurements by Ly et al. (2007) as the area.
Since their way to invoke the proper motion is different from the others and the error is not stated in their papers, we can not put the error bar on those points. 
We consider those fast proper motions as different phenomena (see section 4.3).   
\label{Vapp-z}}
\end{figure*}

\section{Results}

We show the multi-epoch images by EVN observations in Fig. \ref{IPOL-maps}.  
The bright core at the eastern edge of the jet and the continuous jet emission up to 500 mas 
from the core are clearly detected.  The isolated component at the distance of 900 mas from the core corresponds to  
the HST-1 knot.  The jet components between 160 mas to 500 mas have been suggested in previous measurements \citep[]{C07}, 
but they are robustly detected and resolved with high significance in our EVN observations.

In Fig. \ref{IPOL-maps}, green circles mark the model components for the jet structure by fitting circular Gaussians to the emission.  
Red lines represent their individual trajectories over time. We detect no proper motions within the first 160 mas from the core, while we detect significant 
superluminal motions of 2.5 $c$ and 3.5 $c$ for the HST-1. These results are consistent with 
the previous measurements respectively \citep[]{D06, K07, C07}. 
In addition to that, we detected proper motions of the components at the distance between 160 mas and 500 mas from the core for the first time, 
so that the new data allows us to investigate the proper motions of this intermediate region down to HST-1.

Fig. \ref{Vapp-z} shows the apparent velocity ($\beta_{app}$ = $v_{app}$/$c$) for each jet component as a function of 
the projected distance from the core together with previous measurements over the innermost (Reid et al. 1987, Kovalev et al. 2007) 
and outermost regions (Chueng et al. 2007, Biretta et al. 1995; 1999). 
We also show the measurements by \citet[]{L07}, \citet[]{W08} and \citet[]{A09}.  
However, we do not use those measurements in the following discussions, since  errors are not given.  
And we will discuss discrepancy of those measurements in section 4.3.
We identify the smooth transition in the proper motions 
from sub- to super-luminal regimes with a range of 0.67 $c$ $\pm$ 0.63 $c$ to 3.74 $c$ $\pm$ 1.18 $c$ in our EVN observations. 
The apparent velocity increases, 
and connects to the data points in previous measurements at both 160 mas and 900 mas.  
We capture a systematic increase of the apparent proper motions starting from subluminal up to superluminal speeds peaking at around 900 mas, where the HST-1 complex is located.
Beyond the HST-1 complex, the apparent velocities gradually decreases \citep[]{B95, B99, MET13}
Thus, our EVN observations fill the missing link in the velocity distribution of the M87 jet, 
and it enable us to discuss the dynamics of the relativistic jet in this source by compiling observation data over 20 years.

\section{Discussion}

\subsection{Indication of gradual acceleration}

\begin{figure*}
\epsscale{1.1}
\plotone{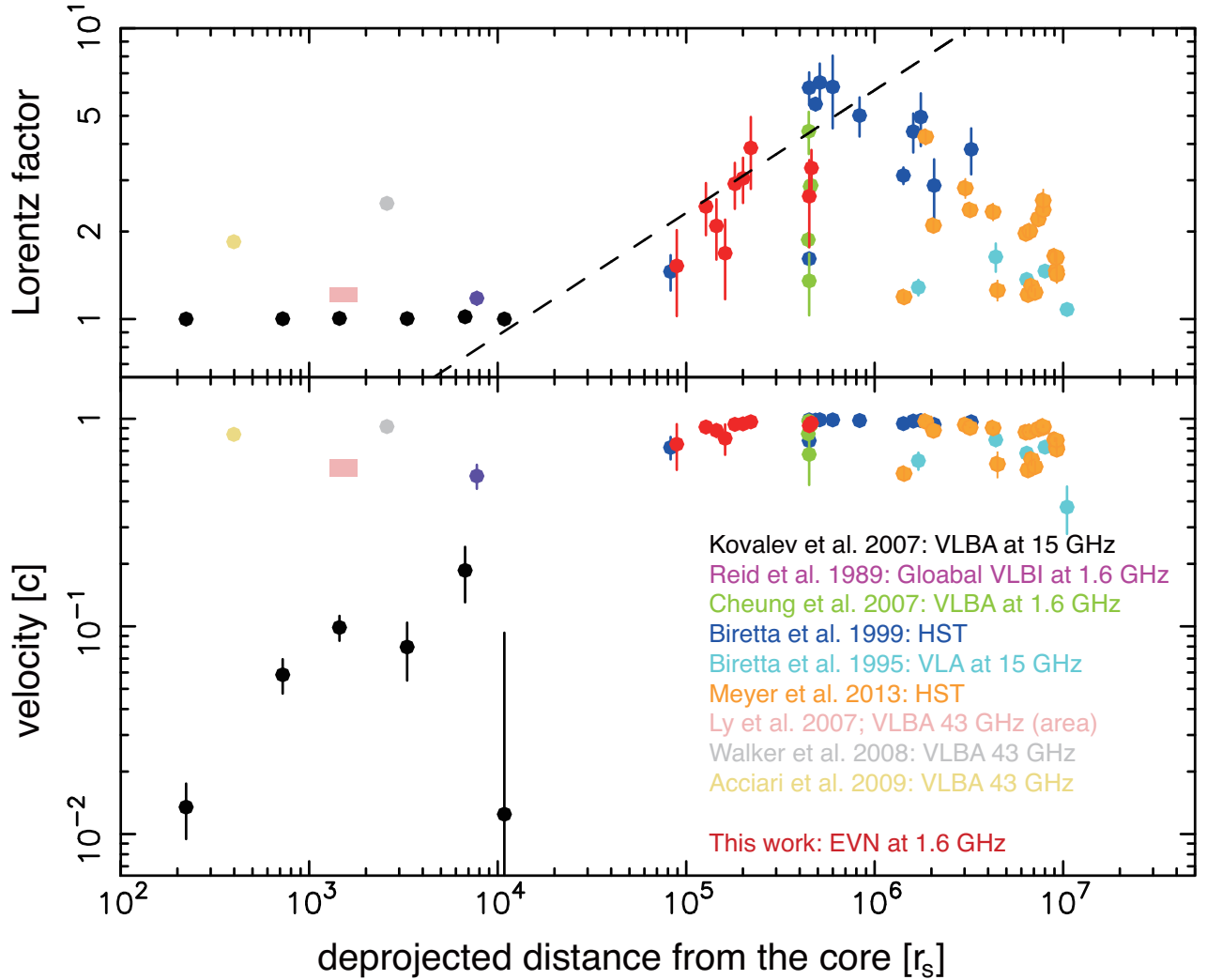}
\caption{Distribution of the velocity and Lorentz factor as a function of the deprojected distance from the core.  
Measurements are illustrated in the same manner as Fig. \ref{Vapp-z}.  
Dashed line present the expected line, $\Gamma \sim z^{(b - 1) / b}$, due to the MHD acceleration in parabolic streamline \citep[e.g.,][]{K09} with its power-law index $b$ = 1.7 \citep[]{AN12}.
Deviation from the predicted line is large at the distance $<$ 10$^{4}$ r$_{s}$, since expected line is applicable for only relativistic regime. \label{Gamma-z}}
\end{figure*}

We show in Fig. \ref{Gamma-z} the velocity and its corresponding Lorentz factor for each jet component as a function of the deprojected distance from the core. 
We assume that a viewing angle of the jet is 14$^{\circ}$ based on the beaming analysis \citep[]{WZ09}. We also assume that it does not change all along the jet.   
We calculate Lorentz factor $\Gamma$=1/(1 - $\beta^{2}$ ) and velocity $\beta$  = $\beta_{app}$/(sin$\theta$ + $\beta_{app}$ cos$\theta$).  

It is clear that both the velocity and the Lorentz factor in the intermediate region of our EVN measurements are connected to the inner and outer regions sampled in previous measurements.  
It is remarkable that the increase of the velocity from non-relativistic (0.01 c) to relativistic (0.97 c) speed takes place over 10$^{2-6}$ $r_{s}$ in the deprojected distance from the core. 
If we simply assume that observed proper motions represent the speed of the bulk flows, 
the gradual acceleration of the bulk flow takes place from 200 $r_{s}$ up to 5 $\times$ 10$^{5}$ 
$r_{s}$ over three orders of magnitude of the distance. 



We note that similar apparent acceleration have been seen in the jet of Cyg A from 0.2 $c$ to 0.7 $c$ at 2 - 10 $\times$ 10$^{3}$ $r_{s}$ \citep[]{K98} and for NGC 315 from 1.1 $c$ to 2.5 $c$ at 5 -20 
$\times$ 10$^{3}$ $r_{s}$ \citep[]{C99}.  However, these acceleration are seen in a limited range of the jets at around 10$^{3}$ r$_{s}$.  
It is probably due to the sensitivity limit, and limited part of the gradual acceleration of the jet are seen for those objects.

\subsection{Comparison with the structure of the jet}

In the general framework on asymptotic evolutions of magnetohydrodynamic (MHD)  jets in parabolic streamlines, both bulk acceleration and collimation take place simultaneously.
Regarding the structure of the M 87 jet, it has been intensively investigated with the VLA/VLBA observations \citep[e.g., ][]{J99}
and revealed that jet changes its opening angle smoothly from 60$^\circ$ at $\sim$ 0.03  pc to smaller than 10$^\circ$ over $\sim$ 10 pc.  
This decrease of the opening angle is considered  collimation \citep[]{J99}.   
Since the collimation continuously takes place up to further distance from the core, MHD process was considered as the collimation mechanism  \citep[]{J99}.
More recently, based on multi frequency VLBI analysis, the M 87 jet is described with parabolic streamline with a power-law index of 1.7; 
the collimation region is identified to be a few up to 10$^{6}$ $r_{s}$ from the core  up to HST-1, which is close to the Bondi radius \citep[]{AN12, NA13, H13}.  

\citet[]{NA13} suggest an acceleration of the M 87 jet in the sub-relativistic regime on 10$^{2-4}$ r$_{s}$ by considering the magnetic nozzle effect.
If this is the case, 
a continuous acceleration zone may presumably expand from a few $r_{s}$ up to 10$^{6}$ $r_{s}$, corresponding to collimation zone in the M 87 jet.  

On the other hand, the newly revealed velocity field by the EVN observations captures a systematic acceleration of the jet in mildly relativistic regime.  
In the framework of MHD jets in a fully relativistic regime, an acceleration of the jet 
can be expressed as a function of the distance $z$, associated with the parabolic stream line with a power-law index $b$, as $\Gamma \propto z^{(b-1)/b}$ \citep[e.g.,][]{KOM07, K09}.
In the case of M 87, the power-law index; $b$, is determined to be 1.7 \citep[]{AN12}, so that we would expect to see the acceleration of the jet as $\Gamma \sim z^{0.42}$.
We show this power-law line in Fig. \ref{Gamma-z} as the reference.  
We do not perform fitting of it, since this equation is good for highly relativistic regime ($\Gamma >> 1$), while the velocity field revealed by EVN observations is mildly relativistic regime.  
However, the measured proper motions show a similar tendency in the function of the distance from the core.

Since observed acceleration and collimation regions are co-spatial and velocity fields are in good agreement with those expected with 
the parabolic MHD jets in both non-relativistic \citep[]{NA13} and relativistic regime (this paper),  
we suggest magnetic acceleration processes play a role in the dynamics of M 87 jet from  0.01 $c$ (at 200 r$_{s})$ up to 0.97 $c$ (at 5 $\times$ 10$^{5}$ $r_{s}$).
This acceleration and collimation zone would correspond to the entire region within the sphere of influence of the supermassive black hole, which is characterized with Bondi radius.



It is suggested that a parabolic streamline is sustained in the M87 jet towards $<$ 10 $r_{s}$ from the central engine \citep[]{NA13}.
If this is the case, similar magnetic acceleration is expected towards more upper stream of the jet, 
so that future monitoring observations with submm VLBI would be important to reveal an initial MHD acceleration together with identifying the footpoint of the jet.  

\subsection{Discrepancy of proper motions among VLBI observations}


There is a large discrepancy in observed proper motions between 15 and 43 GHz at the same region of 1 - 10 mas scales.  
The jet is optically thin in this region at 15 GHz and  widths of the jet are in good agreement with each other at 15 and 43 GHz \citep[]{AN12}, 
so that we would speculate the jet emissions come from an almost same layer.
We re-produced two images obtained with VLBA at 43 and 15 GHz at close epochs in 2007 (Jan. 27 at 43 GHz and Feb. 5th at 15 GHz).  
We compared two images by restoring the synthesized beam at 43 GHz with that at 15 GHz.  
The obtained images are quite similar to each other, and we can easily identify the bright knots in between two images.  
Therefore, we naturally expect to see similar proper motions even at different frequencies.  
However, the VLBA observation at 43 GHz detected superluminal motions up to 2 c, while VLBA and VLBI observations at the other frequencies including 15 GHz 
detected no or only slow sub-luminal motions (see also Fig. 2).  

\citet[]{W08} explained this discrepancy by the difference of the sampling interval of observations.  
The fast motions observed at 43 GHz were detected in only plume-like components with observation interval shorter than 2 weeks. 
\citet[]{W08} pointed out that these plume-like components are short-living, and an observation interval shorter than 2 weeks is required to fully trace these features. 
On the other hand, the sampling interval at 15 GHz is relatively sparse (every 2 -- 8 months, 5 month in median) in comparison with that at 43 GHz (every 5 and 20 days).  
Due to the short life of the components at 43 GHz with superluminal motions, those motions would not be detected by current 15 GHz observations.  

\citet[]{K07} also considered the possibility of the existence of the fast motions based on their 20 years monitoring, 
but conclude that the fast motions are highly unlikely based on 4 aspects even with their sparse sampling interval.
One possible explanation is that the detected fast proper motions are interpreted as the bulk flow velocity \citep[]{W08}, 
while the slow motions reflect standing shocks and/or pattern velocities associated with the instabilities \citep[]{K07, W08}.  
However, it does not explain why we do not see different proper motions, so that observationally it is still puzzling.
We also note that slow proper motions taken by VLBA 15 GHz increases as a function of the distance from the core, 
and thus an interpretation as standing shocks may be inconsistent.

In theoretical aspects, these observed sub-luminal motions can be explained by an MHD bulk velocity of the flow in a sub-relativistic regime \citep[]{NA13}, 
while they have been interpreted during past decades as standing shocks and/or some patterns of plasma instabilities in supersonic jets \citep[]{K07, W08}.
Moreover, newly observed acceleration from sub- to super- luminal motion is in good agreement with the observational evidence of gradual collimation up 
to 5 $\times$ 10$^{5}$ $r_{s}$, therefore it is straightforwards to consider the slow motions as the velocity of the bulk flow.  
However, even if this is the case, an origin of fast proper motions is unclear.  
It can be associated with some pattern velocity, 
or it can also be associated with bulk velocity of the flow by considering multi-layers on the jet structure.  
If the proper motions observed at 43 GHz is associated with the bulk velocity of the flow, 
we can expect the components with faster proper motions at further downstream of the jet accelerated with an MHD nozzle effect.  
However, current VLBI observations at low frequencies have not detected such motions.  
It would be necessary to conduct simultaneous VLBI observations at 15 and 43 GHz to solve this puzzle.

\acknowledgments

We thank P.T.P. Ho for stimulating  discussions. KA thanks member of Greenland Telescope Project for their warm encouragements. 
The European VLBI Network is a joint facility of European, Chinese, South African and other 
radio astronomy institutes funded by their national research councils.

{\it Facilities:} \facility{EVN, MERLIN, VLBA}.

\begin{deluxetable}{lcccccccccccc}
\tabletypesize{\scriptsize}
\tablecaption{Results of the model fitting}
\tablewidth{0pt}
\tablehead{
\colhead{Component} & \multicolumn{4}{c}{2007} & \multicolumn{4}{c}{2008} & \multicolumn{4}{c}{2009}
\\
& \colhead{I} & \colhead{r} & \colhead{P.A.} & \colhead{d} & 
\colhead{I} & \colhead{r} & \colhead{P.A.} & \colhead{d} &  
\colhead{I} & \colhead{r} & \colhead{P.A.} & \colhead{d}
\\
& \colhead{[mJy]} & \colhead{[mas]} & \colhead{[degree]} & \colhead{[mas]} & 
\colhead{[mJy]} & \colhead{[mas]} & \colhead{[degree]} & \colhead{[mas]} & 
\colhead{[mJy]} & \colhead{[mas]} & \colhead{[degree]} & \colhead{[mas]} 
\\
& \colhead{(1)} & \colhead{(2)} & \colhead{(3)} & \colhead{(4)} & 
\colhead{(5)} & \colhead{(6)} & \colhead{(7)} & \colhead{(8)} & 
\colhead{(9)} & \colhead{(10)} & \colhead{(11)} & \colhead{(12)} 
}
\startdata
C7 & 0.071 & 185.2 & -67.0 & 15.0 & 0.050 & 191.6 & -66.9 & 16.8 & 0.047 & 190.3 & -66.9 & 15.7 \\
C6 & 0.047 & 241.2 & -66.8 & 28.7 & 0.044 & 241.3 & -66.6 & 30.2 & 0.050 & 256.2 & -66.5 & 30.7 \\
C5 & 0.039 & 276.1 & -66.9 & 22.1 & 0.031 & 275.8 & -66.7 & 23.9 & 0.024 & 287.5 & -66.8 & 21.9 \\
C4 & 0.060 & 308.0 & -66.3 & 27.2 & 0.046 & 311.3 & -66.3 & 28.4 & 0.056 & 314.7 & -66.6 & 33.1 \\
C3 & 0.026 & 342.0 & -65.9 & 22.8 & 0.033 & 347.8 & -66.1 & 26.1 & 0.034 & 362.0 & -66.4 & 31.2 \\
C2 & 0.062 & 378.1 & -67.4 & 28.2 & 0.057 & 385.3 & -67.8 & 32.2 & 0.040 & 398.8 & -68.4 & 33.1 \\
C1 & 0.029 & 411.8 & -68.5 & 31.5 & 0.026 & 424.0 & -67.9 & 42.0 & 0.016 & 442.4 & -68.8 & 39.0 \\
\enddata
\tablecomments{Columns are as follows: (1). total I flux density (mJy) at 2007, (2) position offset (mas) 
from the core component at 2007, (3) position angle with respect to the core component in degrees, (4) 
size of the fitted Gaussian (mas), (5) - (8) same at 2008, (9) - (12) same at 2009}
\end{deluxetable}


\begin{thebibliography}{}
\bibitem[Acciari et al. (2009)]{A09} Acciari, V. A. et al. 2009, Science, 352, 444
\bibitem[Asada \& Nakamura (2012)]{AN12} Asada, K., \& Nakamura, M.\ 2012, \apjl, 745, L28 
\bibitem[Biretta et al. (1995)]{B95}
Biretta, J.~A., Zhou, F., \& Owen, F.~N. 1995, \apj, 447, 582
\bibitem[Biretta et al. (1999)]{B99}
Biretta, J.~A., Sparks, W.~B., \& Macchetto, F. 1999, \apj, 520, 621
\bibitem[Blandford \& Payne(1982)]{BP82}
Blandford, R.~D., \& Payne, D.~G.\ 1982, \mnras, 199, 883
\bibitem[Blandford \& Znajek(1977)]{BZ77}
Blandford, R.~D., \& Znajek, R.~L.\ 1977, \mnras, 179, 433
\bibitem[Cheung et al. (2007)]{C07}
Cheung, C.~C., Harris, D.~E., \& Stawarz, \L. 2007, \apj, 663, L65
\bibitem[Cotton et al.(1999)]{C99} Cotton, W.~D., Feretti, 
L., Giovannini, G., Lara, L., \& Venturi, T.\ 1999, \apj, 519, 108 
\bibitem[Dodson et al.(2006)]{D06} Dodson, R., Edwards, 
P.~G., \& Hirabayashi, H.\ 2006, \pasj, 58, 243 
\bibitem[Gebhardt et al.(2011)]{G11} Gebhardt, K., Adams,
J., Richstone, D., et al.\ 2011, \apj, 729, 119
\bibitem[Hada et al. (2011)]{H11}
Hada, K., Doi, A., Kino, M., et al.\ 2011, \nat, 477, 185
\bibitem[Hada et al. (2013)]{H13}
Hada, K., Kino, M., Doi, A., et al.\ 2013, \apj, 775, 70
\bibitem[Jord{\'a}n et al.(2005)]{J05} Jord{\'a}n, A.,
C{\^o}t{\'e}, P., Blakeslee, J.~P., et al.\ 2005, \apj, 634, 1002
\bibitem[Junor et al.(1999)]{J99} Junor, W., Biretta, 
J.~A., \& Livio, M.\ 1999, \nat, 401, 891 
\bibitem[Kellermann et al.(2004)]{K04} Kellermann, K.~I., 
Lister, M.~L., Homan, D.~C., et al.\ 2004, \apj, 609, 539 
\bibitem[Komissarov et al. (2007)]{KOM07} 
Komissarov, S. S., Barkov, M. V. Vlahakis, N.,  \&  K\"{o}nigl, A.\ 2007 \mnras, 380, 51
\bibitem[Komissarov et al. (2009)]{K09} 
Komissarov, S. S., Vlahakis, N., K\"{o}nigl, A. \& Barkov, M. V. \ 2009 \mnras, 394, 1182
\bibitem[Kovalev et al.(2007)]{K07} Kovalev, Y.~Y., Lister,
M.~L., Homan, D.~C., \& Kellermann, K.~I.\ 2007, \apjl, 668, L27
\bibitem[Krichbaum et al.(1998)]{K98} Krichbaum, T.~P., Alef, W., Witzel, A., et al.\ 1998, \aap, 329, 873 
\bibitem[Lister et al.(2009)]{L09} Lister, M.~L., Cohen, 
M.~H., Homan, D.~C., et al.\ 2009, \aj, 138, 1874 
\bibitem[Ly et al.(2004)]{L04} Ly, C., Walker, R.~C., 
\& Wrobel, J.~M.\ 2004, \aj, 127, 119 
\bibitem[Ly et al.(2007)]{L07} Ly, C., Walker, R.~C., 
\& Junor, W.\ 2007, \apj, 660, 200 
\bibitem[Macchetto et al.(1997)]{M97} Macchetto, F.,
Marconi, A., Axon, D.~J., et al.\ 1997, \apj, 489, 579
\bibitem[Marshall et al.(2002)]{M02} Marshall, H.~L., 
Miller, B.~P., Davis, D.~S., et al.\ 2002, \apj, 564, 683 
\bibitem[Meier et al.(2001)]{MKU01} Meier, D.~L., Koide, S.,
\& Uchida, Y.\ 2001, Science, 291, 84
\bibitem[Meyer et al.(2013)]{MET13} Meyer, E.~T., Sparks, 
W.~B., Biretta, J.~A., et al.\ 2013, \apjl, 774, L21 
\bibitem[Nakamura \& Asada (2013)]{NA13} Nakamura, M., \& Asada, K. \ 2013, \apj, 775, 118
\bibitem[Owen et al. (1989)]{O89}
Owen, F.~N., Hardee, P.~E., \& Cornwell, T.~J. 1989, \apj, 340, 698
\bibitem[Perlman et al.(1999)]{P99} Perlman, E.~S., 
Biretta, J.~A., Zhou, F., Sparks, W.~B., 
\& Macchetto, F.~D.\ 1999, \aj, 117, 2185 
\bibitem[Reid et al.(1989)]{R89} Reid, M.~J., Biretta,
J.~A., Junor, W., Muxlow, T.~W.~B., \& Spencer, R.~E.\ 1989, \apj, 336, 112
\bibitem[Sparks et al. (1996)]{S96}
Sparks, W.~B., Biretta, J.~A., \& Macchetto, F. 1996, \apj, 473, 254
\bibitem[Walker et al.(2008)]{W08} Walker, R.~C., Ly, C.,
Junor, W., \& Hardee, P.~J.\ 2008, Journal of Physics Conference
Series, 131, 012053
\bibitem[Walsh et al.(2013)]{W05} Walsh, J.~L., Barth, 
A.~J., Ho, L.~C., \& Sarzi, M.\ 2013, \apj, 770, 86 
Biretta, J.~A., Sparks, W.~B., \& Macchetto, F. 1999, \apj, 520, 621
\bibitem[Wang \& Zhou (2009)]{WZ09}
Wang, C.-C., \& Zhou, H.-Y. 2009, \mnras, 395, 301
\end{thebibliography}
\end{document}